# High-speed Low-consumption sEMG-based Transient-state micro-Gesture Recognition

**YOUFANG HAN**, Goertek Inc., China

**WEI ZHAO**, Goertek Inc., China

**XIANGJIN CHEN**, Goertek Inc., China

**XIN MENG***, Goertek Inc., China

*Corresponding author: Xin Meng, shawn.meng@goertek.com, Goertek Inc.

**ABSTRACT**

Gesture recognition on wearable devices is extensively applied in human-computer interaction. Electromyography (EMG) has been used in many gesture recognition systems for its rapid perception of muscle signals. However, analyzing EMG signals on devices, like smart wristbands, usually needs inference models to have high performances, such as low inference latency, low power consumption, and low memory occupation. Therefore, this paper proposes an improved spiking neural network (SNN) to achieve these goals. We propose an adaptive multi-delta coding as a spiking coding method to improve recognition accuracy. We propose two additive solvers for SNN, which can reduce inference energy consumption and amount of parameters significantly, and improve the robustness of temporal differences. In addition, we propose a linear action detection method TAD-LIF, which is suitable for SNNs. TAD-LIF is an improved LIF neuron that can detect transient-state gestures quickly and accurately. We collected two datasets from 20 subjects including 6 micro gestures. The collection devices are two designed lightweight consumer-level sEMG wristbands (3 and 8 electrode channels respectively). Compared to CNN, FCN, and normal SNN-based methods, the proposed SNN has higher recognition accuracy. The accuracy of the proposed SNN is 83.85% and 93.52% on the two datasets respectively. In addition, the inference latency of the proposed SNN is about 1% of CNN, the power consumption is about 0.1% of CNN, and the memory occupation is about 20% of CNN. The proposed methods can be used for precise, high-speed, and low-power micro-gesture recognition tasks, and are suitable for consumer-level intelligent wearable devices, which is a general way to achieve ubiquitous computing.

Additional Key Words and Phrase: Electromyography, spiking neural network, gesture recognition, action detection, wearable device

## 1. INTRODUCTION

In recent years, wrist-worn devices (e.g., smartwatches, smart bands) are becoming mainstream in the field of consumer electronics gradually. The rapid development of worn sensing technology on wearable devices has led to the emergence of many gesture recognition systems [1]. Gesture recognition is an important way to achieve human-computer interaction. It is easy to map human intentions to machine behavior with biosensing technologies and AI technologies. Physical interfaces, like touch screens and keyboards, can be replaced with the conscious control of biological signals by the users. In other words, users can control or interact with devices by non-contact simple gestures, making devices understand our intentions easily and naturally.

Gesture recognition can be applied in many fields, such as extend reality [2], prosthesis [3, 4], sign languages [5], etc. Recently, many studies have focused on gesture recognition tasks by different sensors. IMU-based methods [6,7, 8, 9, 10, 11] can estimate moving joint angles using an accelerometer and gyroscope to capture the dynamic hand gestures. Pressure-based methods [12, 13] using barometric pressure capture the contact pressure profile to enable accurate hand gesture classification. PPG-based methods [14, 15] detect blood volume changes via a pulse oximeter for hand and finger motion detection. Besides, other sensors like acoustic-based methods [16, 17, 18], FMG-based methods [19, 20], and multi-sensors methods [21, 22, 23] have their advantages in different application scenarios respectively.

One more natural way to recognize hand gestures is through electromyography (EMG) [24]. Electromyography is a technique to record electrical activity associated with muscle contraction, which is widely applied in Biotechnology, Biomechanics, and Neuroscience. When muscle generates actions, the brain cortex will propagate an electrical potential(action potential) along the muscle fibers, called motor fiber activation potential (MFAP). The superposition of all MFAP within a motor unit generates a signal called the motor unit action potential (MUAP). EMG can be viewed as the summation of the MUAP [25]. Compared with other sensing technologies, EMG-based technologies can reflect human intent physiologically and realize more natural and fluent interaction with faster response speed.

With the development of deep learning, many works use complex networks, such as MLP and CNN, to analyze EMG signals and build gesture-recognition systems [26, 27, 28]. When deploying complex networks in embedded devices, it usually has issues like high computational complexity, slow inference speed, high energy consumption, and precision loss. In consumer electronics applications, it is necessary to have high accuracy of gesture recognition while having good physical performances, such as low energy consumption (long endurance time), low latency (fast response), and comfortable wearing (lightweight and aesthetically pleasing). However, most EMG-based gesture recognition schemes on wearable devices cannot meet the above requirements simultaneously. Although the generalization and learning principles of artificial neural networks (ANNs) are generated from biological neural networks, their implementation on hardware still cannot achieve the performance of brains in terms of balancing power consumption, efficiency, and accuracy [29].

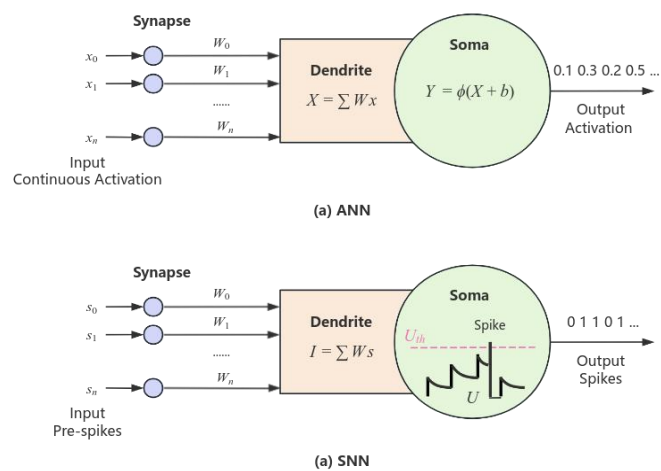

Fig. 1. Comparison of ANN neuron and SNN neuron. They are different in input, output, and Soma function. (a) ANN neuron. Input and output of neuron are continuous activation value. ANN

neuron use non-linear function process real value (b) SNN neuron. Input and output of neuron are Spikes. SNN neuron use LIF function process spikes.

Spiking neural network(SNN) is a third-generation neural network, which is based on the laws of neuromorphic computing [30]. The most important difference between the second-generation network (like MLP) and the third-generation network is the nature of information processing. MLP uses real value, such as signal amplitude, while SNN uses spikes (0 or 1) to process information. Figure 1 compare the neuron in ANN and SNN. SNNs are event-driven methods and are only active when the neuron receives or emits spikes [31]. This characteristic enables SNN to save energy consumption effectively. With the development of neuromorphic hardware, SNNs are easier to develop advantages than ANNs in many fields [32, 33, 34].

This paper proposed a high-speed low-consumption sEMG-based transient-state micro-gesture recognition method by spiking neural network. This method can get high accuracy on transient-state micro-gesture recognition with extremely low power consumption, high inference speed, and low memory occupation. Firstly, the EMG signals have a lower response delay naturally compared with other sensors like IMUs. EMG signals even occur approximately 50-100 ms earlier than the actual action happens [35]. Secondly, SNN-based methods have lower power consumption and faster inference speed compared to traditional ANNs. The proposed method optimizes the network structure to improve the inference speed and recognition accuracy. We also designed an event detection method TAD-LIF for transient-state actions, which can reduce a large amount of computation and locate the target actions quickly. Besides, we developed wristbands with 3 or 8 channels to collect sEMG signals. Compared with Myo [36] and Oymotion [37], our wristbands are lighter, more comfortable, and suitable for daily-wear. The experimental results show that the recognition performance of the proposed method is better than other SNN-based methods, and even better than CNN-based methods on most micro-gestures. The latency of detecting transient-state action is about 50% of baseline methods. The inference latency of the proposed SNN is about 1% of CNN, the power consumption is about 0.1% of CNN, and the memory occupation is about 20% of CNN. The contributions of this paper can be summarized as follows:

- We propose an adaptive multi-delta coding as a spiking coding method, which can improve performance by balancing the average spike rate of input neurons. In the pre-processing stage, we adopt an adaptive normalization to reduce individual differences and adapt different gestures.
- We propose a high-speed transient-state action detection algorithm TAD-LIF, which is suitable for SNN-base methods. TAD-LIF is a modified LIF neuron that can linearly detect target transient-state action from spike trains. TAD-LIF is only active when enough spikes input, which can save energy effectively. We use a Counter to check the detected sequence, excluding the interference from continuous and steady-state actions.
- We propose an SNN-based method for micro-gesture recognition, which has faster inference speed, lower power consumption, and lower memory occupation. We introduce the multi-trains additive solver and multi-steps additive solver into the SNN structure, which can accelerate the inference speed of fully-connected LIF neurons and improve the robustness. In the decision-making part, we adopt population coding to improve the recognition accuracy.
- We collected surface EMG signals of 6 micro-gestures by our designed wristbands. 20

subjects participated in data collection. The experiment results show that the recognition accuracy of the proposed SNN is higher than CNN, FCN, and other SNN-based methods. Compared with CNN, inference latency is about 100 times lower, energy consumption is about 1000 times lower, and the amount of parameters is about 5 times lower.

The remainder of this paper is organized as follows. Section 2 presents the related works of gestures recognition, action event detection, and SNN. Section 3 introduces our wristbands and the description of datasets. Section 4 introduces the proposed method for transient-state micro-gesture recognition, including spike encoding, action detection, and SNN structure. We verify the performance of our method on numerous experiments in Section 5, and discuss the limitations and the future works in Section 6. Finally, we conclude the paper in Section 7.

## 2. RELATED WORK

In this section, we show related works on the EMG-based gesture recognition methods, action event detection methods, and spiking neural network.

### 2.1 EMG-based methods on micro-gestures recognition

Gestures usually have specific symbolic meanings in the HCI (Human Computer Interaction) field. The subtle movements of the fingers are called micro-gestures. Different from steady-state gestures, transient-state micro-gestures have smaller amplitude, faster speed, and lower muscle fatigue impact. If wristbands can recognize mico-gestures of users and transform them into operation instructions, users can interact more naturally with smart devices such as smartphones and computers. Therefore, the accuracy, latency, and consumption of gesture recognition largely determine the practicality and user feeling.

Camera-based gesture recognition systems are sensitive to light and obstacles, IMU-based systems tend to perform well only for overt gestures. In contrast, EMG-based system can solve these issues naturally. EMG signals are transmitted from muscles directly, which contain more subtle movements and force information. According to different collecting methods, EMG signals can be classified into intramuscular EMG signals (iEMG) and surface EMG signals (sEMG). IEMG signals are recorded using needle electrodes planted inside the human muscle, which are unsuitable for smart wristbands. In contrast, sEMG signals are recorded directly on the surface of human skin. Therefore, sEMG is widely used for gesture recognition since it is safe, non-invasive, and easy to implement. According to different electrode numbers, sEMG can be classified into HD-sEMG and LD-sEMG. HD-sEMG is collected by High-density array electrodes, which can capture the distribution of muscle activity with high resolution. However, HD-sEMG usually has a higher cost and cannot be integrated into the smart wristband. LD-sEMG is collected by low-density separate electrodes, which have lower cost, lower consumption on data transmission, and are easy to integrate on wristbands.

At present, most studies use machine learning (ML) and deep learning (DL) techniques for EMG-based gesture recognition. ML-based methods usually calculate the handcrafted features of each electrode signal and input these features into the ML models. Handcrafted features have multi-dimensional information from sEMG signals, like signal amplitude and power (SAP), nonlinear complexity (NLC), frequency information (FI), and time-series features [38]. Then, ML models, such as LDA, SVM, and KNN, are used to analyze these handcrafted features. Chu et al. used LDA on an 8-gesture recognition task, which achieved 97.4% recognition accuracy [39]. Xing et al. used WPT for feature extraction and SVM on a 7-gestures recognition task, which used

4-channels sEMG and achieved 98.4% accuracy [40]. Atlin et al. used KNN on thumb gestures of joystick control and the accuracy was 92% [41].

With the development of DL technology, neural networks can extract implicit information from raw data or handcrafted features. Using a Convolutional Neural Network (CNN) can achieve great gesture recognition performance in user-specific tasks. Triwiyanto et l. classify 10 hand motions from two-channel sEMG by CNN, and the result shows that CNN is better than machine learning algorithms [26]. Ulysses et al. combined CNN and transfer learning to recognize 7-gestures by 8-channels sEMG, which achieved good performance on cross-user applications [27]. Lin et al. proposed DSDAN to achieve unsupervised domain adaptive [28]. They tested DSDAN on HD-sEMG and LD-sEMG, and both got great performance on cross-user gesture recognition. The above works have demonstrated the feasibility of gesture recognition based on sEMG signals.

**2.2 Action event detection**

Action event detection (AED) is used to check whether users perform target gestures. Similar to void activity detection (VAD), micro-gestures can also serve as activation signals for instructions. When users are in a neutral-state or performing non-target actions, AED methods can exclude these irrelative signals and avoid energy waste caused by continuous model inference. The occurrence of gestures always accompanies muscle activity of contraction or extension. Therefore, AED algorithms can utilize the changes in sEMG signals generated by muscle activity to detect the target gestures.

The state-of-the-art AED algorithms can be categorized as visual inspection, statistical methods, machine learning methods, and threshold-based methods [42]. Visual inspection [43] by experts is accurate but complex, and difficult to implement on wristbands. Statistical methods mainly include AGLR [44] and CUSUM [45], which are complex and low-speed. Machine learning methods usually train a discriminator to detect the action [46]. These methods have high detection accuracy, but the latency and power consumption will be significantly increased. Compared to the previous methods, threshold-based methods are found to be the most commonly used on AED for sEMG signals. Threshold-based methods include single threshold, double threshold, and adaptive threshold. Single threshold methods [47] usually compare the difference in adjacent sEMG amplitude. If the difference is over the threshold, this point will be seen as the onset of the action, and then get the whole action segment by a sliding window. Double threshold methods [48] add a second threshold to avoid false positive actions and improve detection accuracy. Adaptive threshold methods [49] segment the signal through SNR or signal energy and adapt different actions with adjustable thresholds.

Threshold-based methods are the simplest and most effective way for AED. However, these methods still have some limitations. 1. The threshold affects the detection performance greatly. 2. These methods are easy to be disturbed by non-target actions. 3. These methods make it difficult to solve the differences in transient-state actions on different subjects with different action durations.

**2.3 Spiking Neural Networks**

SNN is an event-driven method in neuromorphic computing [50]. SNN processes information by a more biologically plausible representation, i.e. spike trains, which have an edge in terms of consumption and latency. Nowadays, SNN is widely applied to object recognition [32], biological signal analysis [33], computer vision [34], etc. Recently, some works have used SNN to implement gesture recognition based on sEMG signals. Sun et al. used SNN on HD-sEMG and LD-sEMG for 9-gesture recognition tasks, which verified the great performance of SNN [51]. SNN not only has faster inference speed, but also relieves the problems of electrode displacement and cross-user difference in sEMG. Xu et al. used SCNN on HD-sEMG for a 6-gesture recognition task, which get a recognition accuracy of 98.78%, even higher than CNN with similar structures [52]. Peng et al. used the NeuCube spiking model on a six-gesture recognition task, which achieved a recognition accuracy of 95.3% [53]. Ma et al. proposed SRNN with STDP and WTA strategies on LD-sEMG for a 5-gesture recognition task [54]. These works demonstrate the feasibility of SNN-based methods on sEMG gesture recognition tasks.

SNN mainly includes four parts: spike encodings, network architectures, training strategies, and hardware realizations. Spike codings are used to convert analog and digital data into spikes [55]. There are a variety of encoding schemes used for SNNs. Two main encoding methods are rate coding and temporal coding [56]. The most commonly used spike encoding method for sEMG signals is delta coding, which is a type of temporal coding.

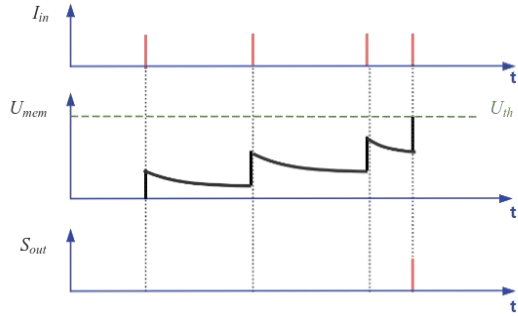

Fig. 2. Change of membrane voltage. $I_{in}$ is input spikes, $U_{mem}$ is membrane voltage, $S_{out}$ is out put spikes. When $U_{mem}$ reaches the threshold $U_{th}$, output spike to next LIF layer.

The basic unit of SNN is the LIF(Leaky Integrate-and-Fire) neuron. LIF simulates the changes in cell membrane voltage, making SNN more biologically plausible. The simplified formula for LIF is shown in Eqs. 1 The three parts in Eqs. 1 correspond to Leaky (voltage decay), Integrate (accumulation of current inputs), and Fire (Fire the spike) respectively.

$$U(t) = \beta U(t-1) + \varepsilon I(t) - S(t-1)U_{th} \qquad (1)$$

Figure 2 shows the change of membrane voltage in LIF. Spike trains are non-differentiable, which makes the training strategies of SNNs different from ANNs. STDP strategy is based on synaptic plasticity, which can be used for unsupervised training [57]. Surrogate gradient descent (SGD) methods are smoothing Dirac functions to make data differentiable by a surrogate function. Besides, other training strategies like STBP [58] and S-TP [59] get good performance in many scenarios.

Finally, SNN is more suitable for realizations on hardware. Traditional ANNs have some limitations on deployment, which heavily rely on computational resources such as GPU or NPU. Most of the ANN-based models choose to deploy in the cloud. In contrast, SNN can be implemented on neuromorphic chips easily with extremely low consumption and high speed. With SNN technologies, it is easier to achieve ubiquitous computing.

## 3. MATERIALS & DATASET

In this section, we introduce our sEMG acquisition system and the datasets used for experiments.

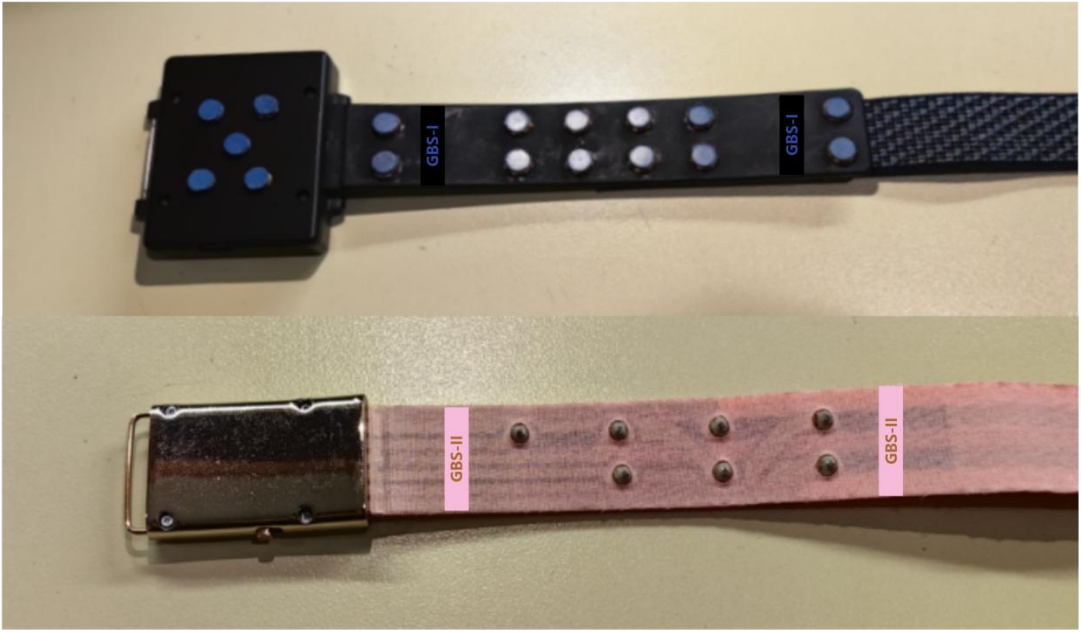

Fig. 3. GBS-I Wristband (top) and GBS-II Wristband (bottom).

### 3.1 sEMG acquisition system

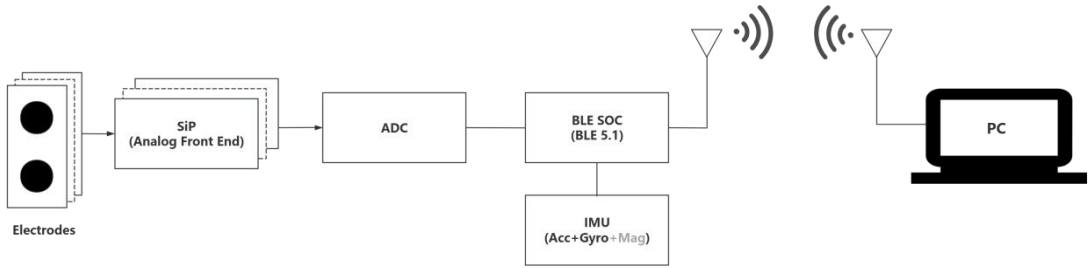

Fig. 4. Schematic diagram of sEMG acquisition system.

We designed and developed two non-invasive devices for recording sEMG at the wrist, named GBS Wristband (Goertek Bio-Sensing), as shown in Figure 3. GBS-I consist an analog front end, an analog-to-digital converter, a buttery, a Bluetooth, an IMU, and electrodes. GBS-I and GBS-II include 8/3 electrodes to record sEMG signals respectively. The sampling rate of two wristbands is 2000 sps, bandwidth is 20-1000 Hz, and BLE 5.1 for data transmission. GBS-I have electrodes made of polymer flexible materials, with 20 $mm^2$ contact dimension and weight of 65 g, and which band is made of silica gel . GBS-II has electrodes made of nanoscale silver paste, with a 7 $mm^2$ contact dimension and a weight of 15 g, and which band is made of flexible material.

Figure 4 shows the schematic diagram of sEMG acquisition system. We use self-developed SiP analog front end for sEMG amplification and filtering. BLE SOC reads the results of ADC, obtains sEMG information in real-time according to 2000Hz per channel, and sends the information to PC in real-time. The specific parameter comparisons with other sEMG acquisition systems are shown in Table 1. Compared with other sEMG acquisition systems, our wristbands are designed to be lighter, more aesthetically pleasing, and more comfortable to wear while ensuring sampling rate and signal quality.

Table 1. Comparison of sEMG acquisition systems.

| | Biometrics DataLITE sEMG | Oymotion gForce-Pro | Thalmic Lab Myo Armband | 3DC Armband | GBS-I | GBS-II |
|---|---|---|---|---|---|---|
| sEMG channels | 16 | 8 | 8 | 10 | 8 | 3 |
| sEMG ADC | 13 bits | 8 bits | 8 bits | 10 bits | 16 bits | 16 bits |
| sEMG sampling rate | 2000 sps | 1000 sps | 200 sps | 1000 sps | 2000 sps | 2000 sps |
| Bandwidth | 10-490 Hz | 20-500 Hz | 5-100 Hz | 20-500 Hz | 20-1000 Hz | 20-1000 Hz |
| Contact dimension | 78 mm$^2$ | ≈66 mm$^2$ | 100 mm$^2$ | 50 mm$^2$ | 20 mm$^2$ | 7 mm$^2$ |
| Contact material | Stainless Steel | Stainless Steel with silver coated | Stainless Steel | Electroless nickel immersion gold | Polymer materials | Silver |
| Input referred-noise | ＜5μV | N.A. | N.A. | 2.2μV | ＜7μV | ＜25μV |
| IMU sensors | / | 9-Axis Acc, Gyro, Mag | 9-Axis Acc, Gyro, Mag | 9-Axis Acc, Gyro, Mag | 6-Axis Acc, Gyro | 9-Axis Acc, Gyro, Mag |
| IMU sampling rate | / | 50 Hz | 50 Hz | 50 Hz | 200 Hz | 200 Hz |
| Transmitter | Wifi | BLE 4.1 | BLE 4.0 | Similar BLE | BLE 5.1 | BLE 5.1 |
| Weight | 17g (per channel) | 80 g | 93 g | 62 g | 65g | 15g |
| Price | ≈$17,000 | $1250 | $200 | ≈$150 | ≈$500 | ≈$300 |

### 3.2 Description of micro-gestures

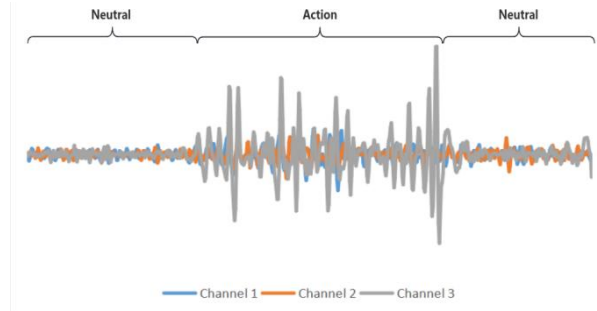

Fig. 5. An example of sEMG signals of transient-state action. A transient-state action include three stages: neutral, action, and neutral.

At present, most studies focus on gesture classification on steady-state gestures, like hand open/close, and finger extension/flexion. For those actions, collected sEMG signals are more stable and can be detected by a window easily. In this paper, we mainly focus on transient-state gestures with faster movement speeds. Compared with normal gestures, micro-gestures have more subtle movement on the fingers. As shown in Figure 4, a complete transient-state micro-gesture includes a process of neutral-action-neutral. When performing a command micro-gesture, the hand will be in a neutral-state first, then make a quick action, and back to a neutral-state in the end.

As shown in Figure 5, we mainly focus on 6 micro-gestures: thumb up, thumb down, thumb left, thumb right, thumb tap, and pinch.

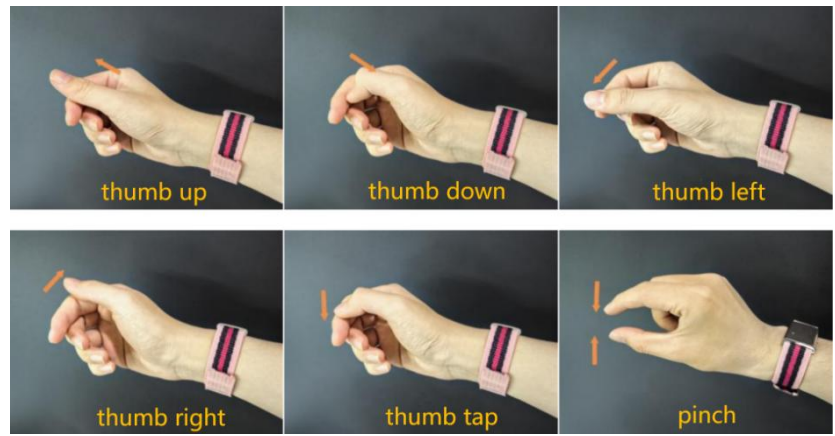

Fig. 6. Sketch map of all micro-gestures. Thumb up/down/left/right means swiping the thumb on the index finger in different orientations. Thumb tap means tapping the thumb on the index finger. Pinch means contacting and releasing the thumb and index finger quickly.

**3.3 Datasets**

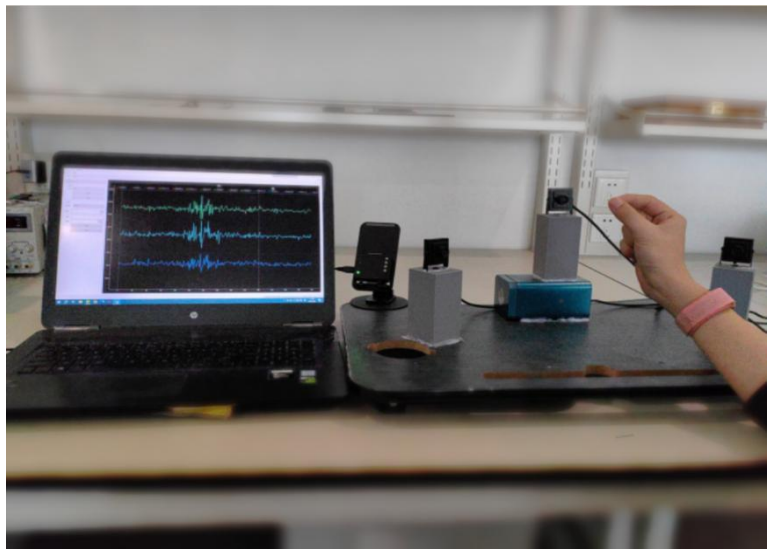

Fig. 7. Overview of sEMG data collection and computer interaction. Collection devices include a PC for guiding and recording the data, a dongle for transmitting data, and three cameras for shooting gesture videos from different angles.

We used two datasets to verify the performance of the proposed method. The dataset I was collected by 8-channels GBS-I, which consists of 7 able-bodied subjects (including 6 males and 1 female, aged 30-50). Dataset I records 5 thumb gestures: thumb up, thumb down, thumb left, thumb right, and thumb tap. Dataset II was collected by 3-channels GBS-II, which consists of 20 able-bodied subjects (including 17 males and 3 females, aged 20-50). Dataset II records pinch and other random gestures.

The data collection process is shown in Figure 7. The collection devices include a PC, a dongle, and three cameras. Each subject sits in front of the collection devices, wearing the GBS wristband and doing gestures under three cameras. The PC will play a guide video first, and then the subject will do corresponding gestures. The sEMG signals collected from the wristband will be transmitted to the PC through Bluetooth, and the PC will process and display the sEMG signal of each channel in real-time. Three cameras will shoot videos from different angles to record the data collection process by the subjects. When problems arise during the subsequent data analysis process, video can be used for tracing.

This study is approved by the Ethics Review Board of the Goertek HCI Research Center (Beijing, China). Informed consent was obtained from all participants.

## 4. METHOD

This section shows our proposed method for transient-state micro-gesture recognition. First, we introduce the spike encoding method on sEMG signal, which transforms the raw data into spike trains. Then, we introduce the TAD-LIF algorithm for transient-state action detection. Finally, we introduce the proposed SNN structure for gesture recognition.

### 4.1 Spike Encoding for sEMG

Before training SNN, it is necessary to convert the raw sEMG signals into spike trains(i.e. binary sequence). Some studies [29, 51, 52] have verified the effectiveness of delta coding methods on sEMG signals. So, we adopt a modified delta coding method as a spike encoding method on sEMG signals. The work of spike encoding and preprocessing includes band-pass filter and rectification, Adaptive normalization, and Adaptive multi-delta coding. To realize better adaptation for users, we collect a piece of neutral-state data and action-state data of each subject at the first use.

#### 4.1.1 Band-pass filter and rectification

Raw sEMG Data usually contains a large amount of noise, so filtering is a necessary step for preprocessing. We use the band-pass Butterworth filter with a bandwidth of 20-500Hz to remove the low-frequency noise and high-frequency interference. Then, calculate absolute values for data to get rectified signals.

#### 4.1.2 Adaptive normalization

After band-pass filter and rectification, the signals are regularized by normalization algorithms. Common normalization algorithms such as the min-max method and z-score method can convert raw data into a fixed range. However, we can not get Maximum, minimum, mean, variance, and other statistics timely while inferencing the input signals online. Besides, due to the existence of individual differences, it is difficult to set suitable statistics in advance to adapt for all users. Therefore, we proposed an adaptive normalization method to solve the above issues.

We first get a segment of neutral-state data as standard data when users first collect sEMG signals. Calculate the median of rectified standard data as M. Then, normalize the data using the following formula.

$$D_{AN} = min(1, max(0, \frac{D_{rec} - M}{\alpha M})) \qquad (2)$$

In this formula, Drec represents rectified data, DAN represents data after adaptive normalization, and $\alpha$ is an adjustable hyperparameter. The median of neutral-state data can reflect the state of sEMG signals of non-action. When collecting data, some slight actions may cause fluctuations in sEMG signals, or other external disturbances may cause abnormally high amplitude that interfere with the results. Therefore, we use the median of the data instead of the mean to reduce the impact of outliers.

### 4.1.3 Adaptive multi-delta coding

Delta coding is a temporal coding method, which can track the temporal changes of the signal amplitude. As shown in Eqs. 3 and Eqs. 4, we can get the spike by the difference of amplitude at adjacent time points. d(t) means signal amplitude at time t, θ is a threshold to determine if the spike occurs. s(t) means whether occurs spike at time t, which takes the value of 0 or 1.

$$\text{diff}(t) = d(t) - d(t-1) \tag{3}$$

$$s(t) = \begin{cases} 0, |\text{diff}(t)| < \theta \\ 1, |\text{diff}(t)| \geq \theta \end{cases} \tag{4}$$

However, there usually have differences in action strength and signal amplitude when performing the same gesture in different subjects. It is difficult to use a fixed threshold θ to adapt all gestures by normal delta coding methods. We proposed an adaptive multi-delta coding method to encode the sEMG signal into a set of spike trains. The pseudo-code of adaptive multi-delta coding is shown in Algorithm 1.

---

**Algorithm 1** Adaptive multi-delta coding

    **Inputs**: sEMG signals
    **Initialization**: minimum threshold $\theta_{min}$, maximum spike rate $R_{max}$, threshold increment δ, number of spike trains N;
    **If** is the **First time**:
        **Calculate** spike trains S by delta coding with $\theta_{min}$, raw sEMG signals → spike trains S
        **While** spike rate of S > $R_{max}$ :
            $\theta_{min}$ += δ;
            **Update** S by delta coding with $\theta_{min}$;
    $\theta = \theta_{min}$
    **For** i in range(1, N) :
        **Calculate** Si by delta coding with $\theta$;
        $\theta$ += δ;
    **Outputs**: {S1, S2, ..., SN}

---

In this algorithm, we set a maximum spike rate to ensure the frequency of the spikes occurring in a small range. We collect a piece of action-state data as a standard to determine the minimum threshold $\theta_{min}$ of delta coding. After that, we adopt N increasing thresholds to generate N spike trains.

### 4.2 TAD-LIF: LIF for transient-state action detection

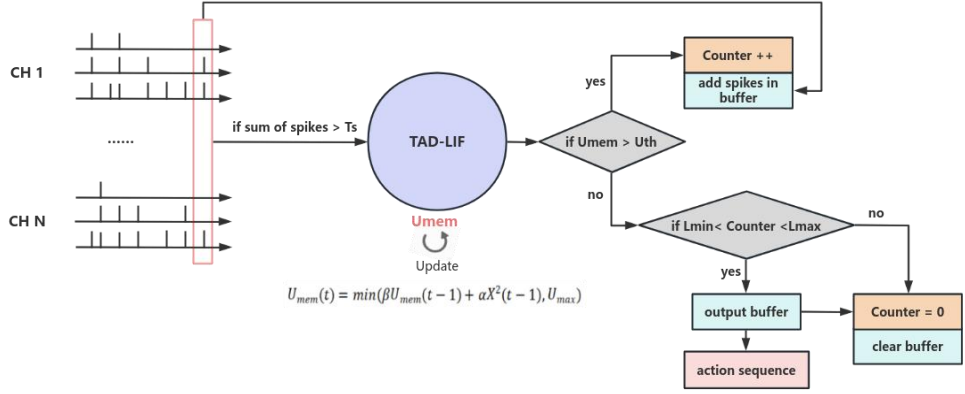

Fig. 8. Overall of TAD-LIF.

Transient-state actions can be seen as instructions that trigger different machine commands to operate smart devices. Different from steady-state action, transient-state action has a shorter duration and usually presents a pattern of *neutral-action-neutral*. Therefore, the goal of TAD-LIF is to detect the occurrence of transient-state actions and extract their signal segments for subsequent model inference.

For digital signals and spike trains, most methods detect action segments by a sliding window with overlap or just input all segments to recognition models directly. These methods may lead to the following issues: 1. the onset of actions is unclear, 2. a large number of invalid calculations, 3. weak ability of anti-interference.

We proposed TAD-LIF to detect transient-state action linearly without windows, as shown in Figure 8. TAD-LIF consists of a modified LIF neuron, an action sequence buffer, and a Counter. TAD-LIF is used to linearly judge whether the spikes are in the action-state, the action sequence buffer is used to store spike trains, and the Counter is used to record the length of the action sequence in the buffer. TAD-LIF neuron uses the following equation to update membrane potential:

$$U_{mem}(t) = min(\beta U_{mem}(t-1) + \omega X^2(t-1), U_{max}) \tag{5}$$

In Eqs. 5, $U_{mem}(t)$ represents the membrane potential at time t, $\beta$ is the decay rate of membrane potential, X(t) represents the sum of spikes on each channel at time t, $\omega$ is the preset weight of each channel. Compared with normal LIF, TAD-LIF does not emit spikes when the $U_{mem}$ reaches the threshold, and the $U_{mem}$ is not reset. Instead, TAD-LIF will send a signal of action-state and add the current spikes into the buffer. $U_{max}$ is an upper limit of the $U_{mem}$, which prevents the potential from endlessly rising and keeps the potential in a reasonable range.

As shown in Figure 8, when the sum of spikes in all channels over threshold Ts, TAD-LIF is activated and updates the $U_{mem}$. As long as the $U_{mem}$ is higher than the threshold $U_{th}$, the signals are considered in the action-state. Then update the Counter and input spikes into the buffer. It is considered as the end of the action when the $U_{mem}$ is lower than the $U_{th}$. We use $L_{min}$ and $L_{max}$ to limit the length of the target action sequence. If the length of the action sequence in the buffer satisfies the requirement of transient-state action, output the action sequence in the buffer to the subsequent SNN model. Otherwise, clear the Counter and the buffer directly.

The advantages of TAD-LIF are summarized as:

1. TAD-LIF can avoid most interference caused by irrelevant movements through controlling the

membrane potential.

2. TAD-LIF can filter out continuous and steady-state actions by Counter.

3. TAD-LIF is a linear algorithm to detect action sequences. TAD-LIF is not activated if no action occurs, so it has higher detection speed and lower computational consumption.

4. Without sliding windows, TAD-LIF does not have a fixed end position. So TAD-LIF can adapt to the action duration of different users to some extent.

## 4.3 Structure of SNN

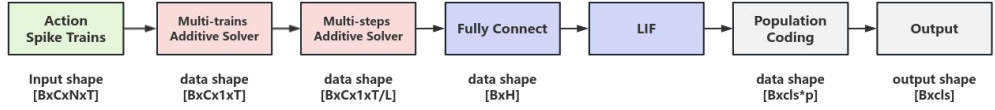

Fig. 9. Structure of proposed SNN.

The structure of the proposed SNN is shown in Figure 9. First, input a set of action spike trains into the network. The input shape is BxCxNxT, B is the batch size (B=1 on test phase), C is the number of channels, N is the number of spike trains, and T is the action length. The input data will be processed by two additive solvers. Multi-trains additive solver will add up data on the dimension of N. The data shape is transformed from BxCxNxT to BxCx1xT. We can obtain the total spike intensity on each time step by this solver. A multi-step additive solver will segment and add up data by a window L. The data shape is transformed from BxCx1xT to BxCx1xT/L. This additive solver splits the action sequence into many smaller action segments, which can not only reduce data size, but also improve the robustness to time-difference of actions.

After two additive solvers, flatten the data to the shape of BxH (H=C*T/L). Then, input the data to a Fully Connect layer and a normal LIF layer. Eqs. 6 shows the change of membrane potential in normal LIF. Umem(t) represents the membrane potential at time t, $\beta$ is the decay rate of membrane potential, X(t) represents the spikes emitted from pre-layers, $\omega$ is the weight of each neuron, S(t) represents the occurrence of spikes (0 or 1) at time t, $U_{th}$ is the threshold to control spike emission. In the output layer, we use population coding to improve the recognition performance. The population coding method assigns p output neurons to each class instead of one neuron per class. The final output gesture class is determined by the summation of p neurons of each class, which is shown in Eqs. 7.

$$U_{mem}(t) = \beta U_{mem}(t-1) + \omega X(t) - S(t-1)U_{th} \qquad (6)$$

$$\text{Output} = \text{argmax}(\sum S_{cls}) \qquad (7)$$

Spiking function S(t) and its derivative function are shown in Eqs. 8 and Eqs. 9. Θ(x) is the Heaviside function, when x>0, the result is 1, otherwise the result is 0. δ(x) is the Dirac function, when x ≠ 0, the result is 0, otherwise the result is infinite.

$$S(t) = \Theta(U(t) - U_{thr}); \quad \Theta(x) = \begin{cases} 0, x \leq 0, \\ 1, x > 0, \end{cases} \qquad (8)$$

$$\frac{\partial S}{\partial U} = \delta(U - U_{thr}); \quad \delta(x) = \begin{cases} 0, x \neq 0, \\ \infty, x = 0, \end{cases} \qquad (9)$$

The weight of LIF can not be updated by the gradient descent method directly, because spike signals are Heaviside functions, whose derivative δ(x) is non-differentiable. We use the surrogate gradient descent method to solve this issue. Fast sigmoid can be used as a smooth function of the Dirac function, which is differentiable. Fast sigmoid and its derivative are shown in Eqs. 10 and

Eqs. 11. In the formulas, k is a hyperparameter to adjust the smoothness extent. When k→∞, $\frac{\partial \tilde{S}}{\partial U} = \delta(U - U_{thr})$.

$$\tilde{S}(t) = \frac{U(t) - U_{thr}}{1 + k|U(t) - U_{thr}|} \quad (10)$$

$$\frac{\partial \tilde{S}}{\partial U} = \frac{1}{(k|U - U_{thr}| + 1)^2} \quad (11)$$

## 5. EXPERIMENTS

We conducted experiments on two datasets, verifying the performance of proposed TAD-LIF and SNN methods.

### 5.1 Performance of TAD-LIF

#### 5.1.1 Experiment setting

TAD-LIF has lots of hyper-parameters. The hyperparameter settings in this experiment are as follows: Threshold Ts is 5, weight α is 1, decay rate β is 0.95, and maximum membrane potential Umax is 5.

We compared TAD-LIF with the non-spike threshold-based method and the spike threshold-based method. Non-spike threshold-based method uses a 300ms window with 50% overlap, and compares the difference of maximum and median of rectified signals in this window. If the difference is over than a fixed threshold, signals in the windows are recognized as the target segment. The spike threshold-based method sets an onset when the sum of spikes at time t is over the threshold Ts. If the sum of spike trains in a 300ms window is over a fixed threshold, these spike trains are recognized as the target segment.

We verify the performance of the algorithms by metrics of Recall, Precision, and Latency. Recall represents the ratio of correctly detected target segments, Precision represents the ratio of the target segments in all detected actions. Therefore, high Recall means more target segments can be found, while high Precision means less inference burden of subsequent models.

#### 5.1.2 Recall & Precision

Table 2. Recall and Precision of different methods. For each Dataset, the optimal results are **bold**, and the sub-optimal results are underlined. TAD-LIF achieves the best Precision and the second best Recall on two datasets.

| Method | Dataset I | | Dataset II | |
|---|---|---|---|---|
| | Recall/% | Precision/% | Recall/% | Precision/% |
| Non-spike Threshold | **99.33** | 10.24 | **99.84** | 7.89 |
| Spike Threshold | 91.85 | 3.09 | 97.60 | 4.95 |
| **TAD-LIF** | 97.67 | **27.76** | 99.68 | **56.12** |

The performance of three algorithms on Dataset I and Dataset II are shown in Table 2. Non-spike threshold-based method have the best Recall, but Precision is much lower than TAD-LIF. Low precision will cause more irrelevant action sequences to be input into the subsequent model, resulting in a large inference burden. The spike threshold-based method and TAD-LIF use spike trains to detect target sequences. Compared with the spike threshold-based method with sliding window, TAD-LIF has higher Recall and Precision and is more suitable for spike trains.

### 5.1.3 Latency

Spikes are emitted only when actions occur, i.e. most of the neutral-state have no spikes. Therefore, there is no need for action detection in most neutral-state. To compare the delay of three algorithms, we experiment with different action rates. We set the 100% action rate as 100 gestures occur within 100 seconds. The results are shown in Table 3 and Figure 10. In Table 3, the latency of TAD-LIF is about 50% of threshold-based methods. In Figure 10, the latency of the non-spike method shows an irregular change pattern, while spike-based methods show that the delay rises with the increase of action rate. Therefore, TAD-LIF has a superior advantage with a low action rate.

Table 3. Latency of different methods. For each action rate, the lowest latency is **bold**. TAD-LIF has the lowest latency in all action rates.

| Action Rate | Non-spike Threshold/s | Spike Threshold/s | **TAD-LIF/s** |
|---|---|---|---|
| 2% | 3.11 | 2.68 | **1.16** |
| 10% | 2.88 | 2.78 | **1.35** |
| 25% | 3.22 | 2.87 | **1.48** |
| 50% | 3.19 | 2.99 | **1.59** |
| 75% | 3.14 | 3.29 | **1.77** |
| 100% | 2.73 | 3.79 | **2.31** |

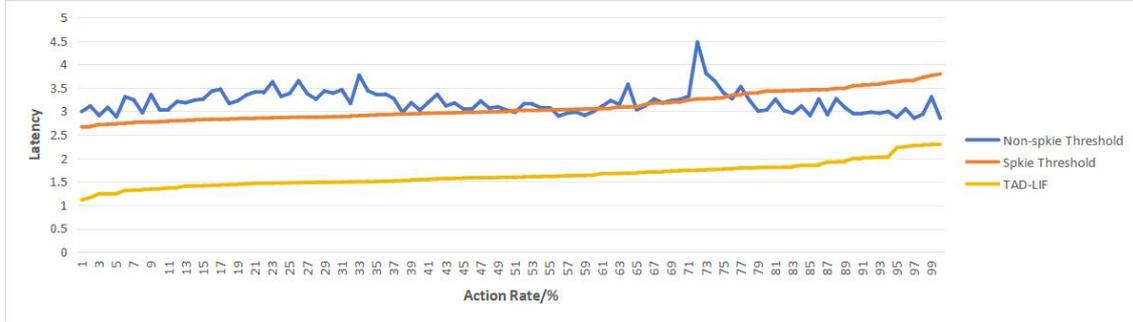

Fig. 10. Latency on different action rates of different methods.

### 5. 2 Performance of proposed SNN
### 5.2.1 Experiment setting

The hyperparameter settings in this experiment are as follows: number of spike trains N in Adaptive multi-delta coding is 10, window length L in Multi-steps additive solver is 20, decay rate β is 0.9, number of output neurons p in population coding is 100.

We compare our method with 4 baseline models. *CNN(Convolutional Neural Network)*, using two 1*5 kernel Convolutional layers with maxpool and Batch Normalization, following a Fully-Connect Layer(FC). *FCN(Fully-Connect Network)*, using two FC layers directly. CNN and FCN use the min-max standardization method as pre-processing. *SCNN(Spiking Convolutional Neural Network)*, uses the same structure as CNN and replaces the activation function with LIF. *Normal SNN*, using the same structure as FCN replaces the activation function to LIF. SCNN uses rate coding while normal SNN uses delta coding as pre-processing. The proposed SNN has same LIF layers as normal SNN.

In this part, we use accuracy, latency, power consumption, and memory occupation as evaluating indicators.

**5.2.2 Accuracy**

We compared the accuracy of the proposed SNN with CNN, FCN, SCNN, and normal SNN on Dataset I and Dataset II. Each experiment was repeated 10 times with different random seeds, and the mean and the standard deviation were recorded. The results are shown in Table 4 and Figure 11. Dataset I includes 5 thumb-related gestures, while Dataset II includes pinch and other gestures. In Table 4, the proposed SNN gets the best accuracy of 83.85% on Dataset I and 93.85% on Dataset II. Which accuracy is extremely higher than other SNN-based methods, even higher than CNN. A one-way repeated-measure ANOVA was applied to the accuracy of the proposed SNN versus the other contrast methods. The ANOVA result shows proposed SNN had significantly better performance than other SNN-based methods. Figure 11 shows the accuracy of each gesture on 5 methods. FCN, SCNN, and normal SNN have large variances in each gesture and between gestures. The proposed SNN shows better stability.

Table 4. Accuracy of different methods for all subjects. For each Dataset, the optimal results are **bold**, and the sub-optimal results are underlined. Left of $\pm$ is the mean and right of $\pm$ is the standard deviation. The value in () means significant level p. Our proposed SNN gets the best accuracy on two datasets.

| Method | *Dataset I/%* | *Dataset II/%* |
|---|---|---|
| CNN | 81.99±0.65 (0.205) | 92.83±0.52 (0.811) |
| FCN | 64.10±1.30 (<0.001**) | 92.34±0.42 (0.389) |
| SCNN | 60.18±3.04 (<0.001**) | 84.18±2.93 (<0.001**) |
| Normal SNN | 55.76±2.51 (<0.001**) | 87.51±1.19 (<0.001**) |
| Proposed SNN | **83.85±0.63** | **93.52±0.43** |

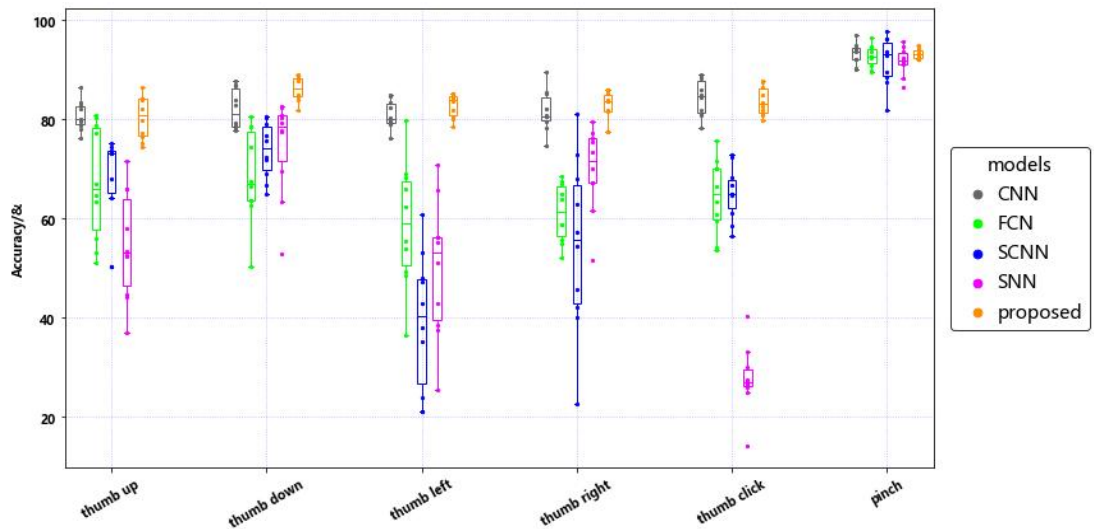

Fig. 11. The box-plot of recognition accuracy on 6 gestures for all subjects. The results of thumb-related gestures are tested on dataset I, and the result of pinch gesture is tested on dataset II

The sEMG signals show significant differences in different subjects, so we complement the experiment on one-subject data. The results are shown in Table 5 and Figure 12. In table 5, the proposed SNN can get an accuracy of 95.17% on Dataset I, which is slightly lower than CNN but much higher than normal SNN-based methods. The proposed SNN gets the best accuracy of 98.23% on Dataset II. The ANOVA result shows proposed SNN had significantly better performance than other methods except CNN.

Table 5. Accuracy of different methods for one subject. For each Dataset, the optimal results are **bold**, and the sub-optimal results are underlined. Left of $\pm$ is the mean and right of $\pm$ is the standard deviation. The value in () means significant level p. Our proposed SNN gets the best accuracy on Dataset II and the second best accuracy on Dataset I.

| Method | Dataset I/% | Dataset II/% |
|---|---|---|
| CNN | **97.73$\pm$0.39 (<0.01*)** | 96.95$\pm$0.91 (0.196) |
| FCN | 92.43$\pm$0.73 (<0.001**) | 96.36$\pm$1.20 (0.019) |
| SCNN | 87.54$\pm$1.40 (<0.001**) | 96.92$\pm$1.40 (0.177) |
| Normal SNN | 80.36$\pm$2.45 (<0.001**) | 96.35$\pm$1.50 (0.018) |
| Proposed SNN | 95.17$\pm$0.82 | **98.23$\pm$1.39** |

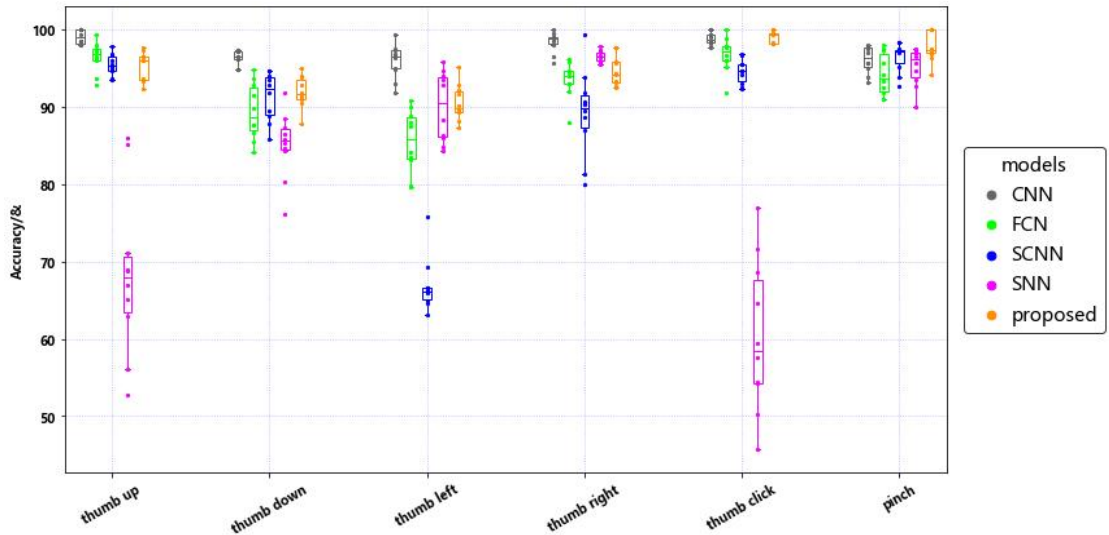

Fig. 12. The box-plot of recognition accuracy on 6 gestures for one subject. The results of thumb-related gestures are tested on dataset I, and the result of pinch gesture is tested on dataset II

As shown in Figure 12, the prediction of each gesture is very unstable by using normal SNN or SCNN directly. With the same LIF layers as normal SNN, the proposed SNN can effectively improve stability and accuracy.

**5.2.3 Model physical performance**
In this part, we compared the Inference latency, power consumption, and memory occupation of different methods. Inference latency is the time required to infer one sample. Power consumption is calculated by the number of accumulation (AC) and multiply-accumulation (MAC). Horowitz et al. showed that 32-bit integer AC consumes about 0.1pJ while MAC consumes about

3.2pJ on 45nm CMOS25 [60]. According to the results of Horowitz, we estimate the power consumption of inferring one sample on different models. CNN mainly calculates the consumption of convolutional layers and fully-connect layers. FCN mainly calculates the consumption of fully-connect layers. SNN-based methods calculate the consumption of convolutional/fully-connect layers, LIF layers, and spike encoding. Amount of parameters is the number of trainable parameters required for model inference, which represents the memory occupation.

Table 6. Physical performance of different methods. The optimal performances are **bold**, and the sub-optimal performances are underlined. SNNs are better than other methods. Our proposed SNN gets the best performances.

| Method | Inference latency/s | power consumption/pJ | Amount of parameters |
|---|---|---|---|
| CNN | $4.89 \times 10^{-3}$ | $\approx 2.96 \times 10^7$ | 208066 |
| FCN | $3.58 \times 10^{-4}$ | $\approx 3.60 \times 10^5$ | 126274 |
| SCNN | $3.70 \times 10^{-4}$ | $\approx 1.09 \times 10^6$ | 207874 |
| Normal SNN | $\underline{6.31 \times 10^{-5}}$ | $\approx \underline{1.21 \times 10^5}$ | $\underline{115586}$ |
| Proposed SNN | $\mathbf{3.12 \times 10^{-5}}$ | $\approx \mathbf{2.83 \times 10^4}$ | **44588** |

The results in Table 6 are calculated from Dataset II. Models are implemented on CPU (Intel Core i7-6700HQ@2.60GHz). The results show that the proposed method has the best performance on inference latency, power consumption, and memory occupation. The power consumption of the proposed SNN is extremely lower than CNN. This is because the ANNs require a large amount of multiplication operations, while SNN-based methods can convert many multiplication into addition operations. SNN-based methods only have multiplication in updating membrane potentials. Besides, the proposed SNN adds two additive solvers before LIF layers, which further compress the data scale, resulting in lower energy consumption and fewer amount of parameters than normal SNN.

**5.3 Ablation experiment**

This part verified the effectiveness of each component of the proposed SNN by an ablation experiment. The results are shown in Table 7 and Figure 13. In Table 7, the first two lines are the results of substituting adaptive multi-delta coding with normal delta coding and rate coding respectively. From the results, changing the spike coding method may significantly decrease the prediction accuracy, especially using normal delta coding directly. Without population coding or additive solvers, the prediction accuracy of most gestures also decreases significantly, especially thumb-related gestures. From the ANOVA result, change or remove the component of proposed method will decrease the performance significantly. Therefore, adaptive multi-delta coding, population coding, and additive solvers are indispensable components of the proposed SNN.

Table 7. Accuracy of ablation experiments. For each Dataset, the optimal results are **bold**, and the sub-optimal results are underlined. Left of $\pm$ is the mean and right of $\pm$ is the standard deviation. The value in () means significant level p. Change or remove the component of proposed method will decrease the accuracy on two datasets.

| Method | *Dataset I/%* | *Dataset II/%* |
|---|---|---|
| normal delta coding | 59.75±1.87 (<0.001**) | 91.76±0.53 (<0.001**) |
| rate coding | 78.00±1.23 (<0.001**) | 92.68±0.35 (<0.001**) |
| w.o. population coding | 69.77±1.44 (<0.001**) | 92.14±0.47 (<0.001**) |
| w.o. additive solvers | 64.96±1.92 (<0.001**) | 86.23±0.67 (<0.001**) |
| proposed | **83.85±0.63** | **93.52±0.43** |

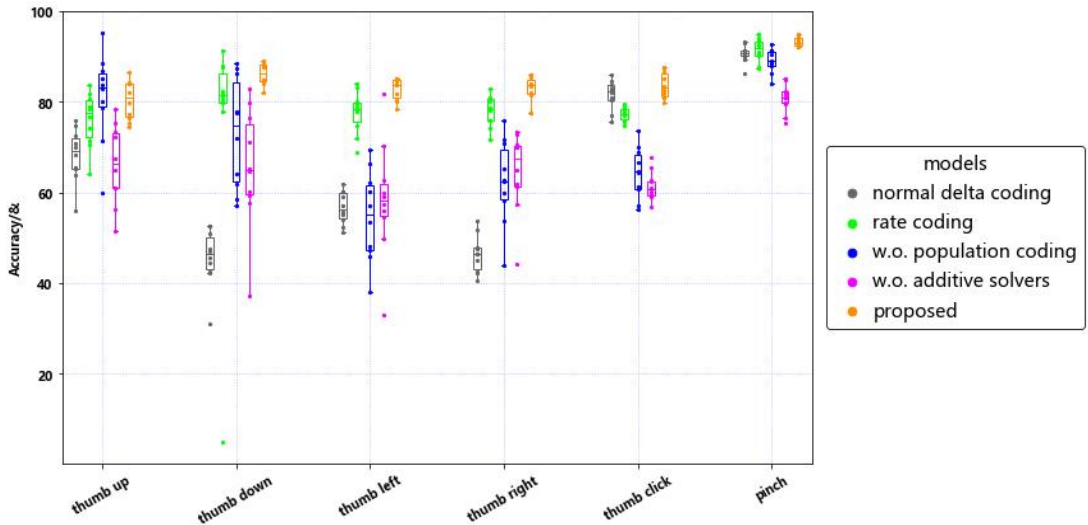

Fig. 13. The box-plot of recognition accuracy on 6 gestures in ablation experiment. The results of thumb-related gestures are tested on dataset I, and the result of pinch gesture is tested on dataset II

## 6. DISCUSSION

In this paper, we have proposed a method for micro-gestures recognition with high accuracy, high speed, and low consumption. In the following, we discuss the limitations and the future work of our study.

**6.1 Limitation**

There are still some limitations in our works. Firstly, our method is still difficult to distinguish some similar micro-gestures. On the one hand, LD-sEMG has a lower resolution than HD-sEMG, so it is difficult to extract accurate features to distinguish some similar gestures. On the other hand, the structure of the proposed SNN is simple for getting better physical performance, so it is hard to extract deeper implicit features. Secondly, we did not consider the cross-user scenarios. The significant difference in different users is the biggest issue of sEMG. Many works adopt transfer learning methods to address the issue of cross-users [61, 62]. Some works verified that SNN itself can relieve the influence of electrode shift and cross-user difference to some extent [51, 52], but the performances are not as good as transfer learning methods. Finally, our experiments are implemented on the CPU now. We have not tested our method on SNN-based hardware, like neuromorphic chips, so we are not sure if it can reach the same performance on embedded devices.

**6.2 Future work**

We will implement more future works on the aforementioned limitations. We will do more experiments to find a balance in deepening network structure for high recognition accuracy and keeping physical performances. We will collect more data on other micro-gestures from more subjects to verify the generalization capability. We will further investigate how to combine the proposed SNN with transfer learning tricks to address the issue of cross-users. Besides, we will try to implement our SNN method in neuromorphic hardware and verify its performance in wearable devices.

## 7. CONCLUSION

To realize gesture recognition on wearable devices better, this paper proposed a novel SNN method with extremely low power consumption, low inference latency, and high recognition accuracy. The proposed method is suitable for transient-state micro-gesture recognition on LD-sEMG. The main contribution of the proposed method is as follows: 1. We use adaptive normalization to reduce the influence of individual differences. 2. We propose an adaptive multi-delta coding, which has better performance than other normal spiking coding methods. 3. We propose TAD-LIF to detect transient-state actions, which have higher speed and precision than other methods. 4. We propose an SNN network with two additive solvers to reduce the scale of the model and improve the robustness to the time-difference of different actions. Besides, we use population coding to improve the prediction accuracy.

We use our designed lighter and more comfortable wristbands to collect sEMG signals and implement experiments. From the experimental results, the proposed SNN has better recognition performance than CNN on most micro-gestures. The inference latency of the proposed SNN is about 1% of CNN, the power consumption is about 0.1% of CNN, and the memory occupation is about 20% of CNN. In conclusion, the proposed method has both better prediction accuracy and better physical performance. Our work indicates that SNN methods are suitable for fast inference and low-consumption scenarios, like applications on embedded devices. This can be a general way to realize the goal of "ubiquitous computing".

## ACKNOWLEDGMENTS
This work was supported by Goertek Institute of Technology and Beijing Goertek Technology Co.,Ltd. We thank the Goertek HCI Research Center for gathering the micro-gestures datasets analyzed in this paper.

## REFERENCES
[1] Zhang M, Dai Q, Yang P, et al. idial: Enabling a virtual dial plate on the hand back for around-device interaction[J]. Proceedings of the ACM on Interactive, Mobile, Wearable and Ubiquitous Technologies, 2018, 2(1): 1-20.
[2] Fashimpaur J, Karlson A, Jonker T R, et al. Investigating Wrist Deflection Scrolling Techniques for Extended Reality[C]//Proceedings of the 2023 CHI Conference on Human Factors in Computing Systems. 2023: 1-16.
[3] Lee S, Sung M, Choi Y. Wearable fabric sensor for controlling myoelectric hand prosthesis via classification of foot postures[J]. Smart Materials and Structures, 2020, 29(3): 035004.
[4] Meng L, Jiang X, Liu X, et al. User-tailored hand gesture recognition system for wearable prosthesis and armband based on surface electromyogram[J]. IEEE Transactions on Instrumentation and Measurement, 2022, 71: 1-16.
[5] Ahmed M A, Zaidan B B, Zaidan A A, et al. A review on systems-based sensory gloves for sign language recognition state of the art between 2007 and 2017[J]. Sensors, 2018, 18(7): 2208.
[6] Zhang C, Bedri A K, Reyes G, et al. TapSkin: Recognizing on-skin input for

smartwatches[C]//Proceedings of the 2016 ACM International Conference on Interactive Surfaces and Spaces. 2016: 13-22.

[7] Chen W, Wang Z, Quan P, et al. Robust Finger Interactions with COTS Smartwatches via Unsupervised Siamese Adaptation[C]//Proceedings of the 36th Annual ACM Symposium on User Interface Software and Technology. 2023: 1-14.

[8] Xu X, Gong J, Brum C, et al. Enabling hand gesture customization on wrist-worn devices[C]//Proceedings of the 2022 CHI Conference on Human Factors in Computing Systems. 2022: 1-19.

[9] Lu Y, Huang B, Yu C, et al. Designing and evaluating hand-to-hand gestures with dual commodity wrist-worn devices[J]. Proceedings of the ACM on Interactive, Mobile, Wearable and Ubiquitous Technologies, 2020, 4(1): 1-27.

[10] Kim M, Cho J, Lee S, et al. IMU sensor-based hand gesture recognition for human-machine interfaces[J]. Sensors, 2019, 19(18): 3827.

[11] Kang P, Li J, Fan B, et al. Wrist-worn hand gesture recognition while walking via transfer learning[J]. IEEE Journal of Biomedical and Health Informatics, 2021, 26(3): 952-961.

[12] Jung P G, Lim G, Kim S, et al. A wearable gesture recognition device for detecting muscular activities based on air-pressure sensors[J]. IEEE Transactions on Industrial Informatics, 2015, 11(2): 485-494.

[13] Shull P B, Jiang S, Zhu Y, et al. Hand gesture recognition and finger angle estimation via wrist-worn modified barometric pressure sensing[J]. IEEE Transactions on Neural Systems and Rehabilitation Engineering, 2019, 27(4): 724-732.

[14] Zhao T, Liu J, Wang Y, et al. Towards low-cost sign language gesture recognition leveraging wearables[J]. IEEE Transactions on Mobile Computing, 2019, 20(4): 1685-1701.

[15] Fang C, Ruan X, Zhang X, et al. Exploration on the Negative Effects of Sensor Shifts in Photoplethysmography-based Gesture Recognition and a Solution based on Transfer Learning[J]. IEEE Transactions on Instrumentation and Measurement, 2023.

[16] Zhang C, Xue Q, Waghmare A, et al. FingerPing: Recognizing fine-grained hand poses using active acoustic on-body sensing[C]//Proceedings of the 2018 CHI Conference on Human Factors in Computing Systems. 2018: 1-10.

[17] Iravantchi Y, Goel M, Harrison C. BeamBand: Hand gesture sensing with ultrasonic beamforming[C]//Proceedings of the 2019 CHI Conference on Human Factors in Computing Systems. 2019: 1-10.

[18] McIntosh J, Marzo A, Fraser M, et al. Echoflex: Hand gesture recognition using ultrasound imaging[C]//Proceedings of the 2017 CHI Conference on Human Factors in Computing Systems. 2017: 1923-1934.

[19] Xiao Z G, Menon C. A review of force myography research and development[J]. Sensors, 2019, 19(20): 4557.

[20] Jiang X, Merhi L K, Xiao Z G, et al. Exploration of force myography and surface electromyography in hand gesture classification[J]. Medical engineering & physics, 2017, 41: 63-73.

[21] Zhao T, Liu J, Wang Y, et al. Towards low-cost sign language gesture recognition leveraging wearables[J]. IEEE Transactions on Mobile Computing, 2019, 20(4): 1685-1701.

[22] Jiang S, Lv B, Guo W, et al. Feasibility of wrist-worn, real-time hand, and surface gesture recognition via sEMG and IMU sensing[J]. IEEE Transactions on Industrial Informatics, 2017, 14(8):

3376-3385.

[23] Wang H, Kang P, Gao Q, et al. A novel PPG-FMG-ACC wristband for hand gesture recognition[J]. IEEE journal of biomedical and health informatics, 2022, 26(10): 5097-5108.

[24] Allard U C, Nougarou F, Fall C L, et al. A convolutional neural network for robotic arm guidance using sEMG based frequency-features[C]//2016 IEEE/RSJ International Conference on Intelligent Robots and Systems (IROS). IEEE, 2016: 2464-2470.

[25] Liu Y, Zhang S, Gowda M. NeuroPose: 3D hand pose tracking using EMG wearables[C]//Proceedings of the Web Conference 2021. 2021: 1471-1482.

[26] Triwiyanto T, Pawana I P A, Purnomo M H. An improved performance of deep learning based on convolution neural network to classify the hand motion by evaluating hyper parameter[J]. IEEE Transactions on Neural Systems and Rehabilitation Engineering, 2020, 28(7): 1678-1688.

[27] Côté-Allard U, Fall C L, Drouin A, et al. Deep learning for electromyographic hand gesture signal classification using transfer learning[J]. IEEE transactions on neural systems and rehabilitation engineering, 2019, 27(4): 760-771.

[28] Lin X, Zhang X, Zhang X, et al. DSDAN: Dual-Step Domain Adaptation Network Based on Bidirectional Knowledge Distillation for Cross-User Myoelectric Pattern Recognition[J]. IEEE Sensors Journal, 2023.

[29] Ma Y, Chen B, Ren P, et al. EMG-based gestures classification using a mixed-signal neuromorphic processing system[J]. IEEE Journal on Emerging and Selected Topics in Circuits and Systems, 2020, 10(4): 578-587.

[30] Izhikevich E M. Polychronization: computation with spikes[J]. Neural computation, 2006, 18(2): 245-282.

[31] Roy K, Jaiswal A, Panda P. Towards spike-based machine intelligence with neuromorphic computing[J]. Nature, 2019, 575(7784): 607-617.

[32] Kim S, Park S, Na B, et al. Spiking-yolo: spiking neural network for energy-efficient object detection[C]//Proceedings of the AAAI conference on artificial intelligence. 2020, 34(07): 11270-11277.

[33] Feng Y, Geng S, Chu J, et al. Building and training a deep spiking neural network for ECG classification[J]. Biomedical Signal Processing and Control, 2022, 77: 103749.

[34] Zhou C, Yu L, Zhou Z, et al. Spikingformer: Spike-driven Residual Learning for Transformer-based Spiking Neural Network[J]. arXiv preprint arXiv:2304.11954, 2023.

[35] Artemiadis P. EMG-based robot control interfaces: past, present and future[J]. Advances in Robotics & Automation, 2012, 1(2): 1-3.

[36] Thalmics Labs Myo Armband. Available online: https://support.getmyo.com/hc/en-us.

[37] Oymotion G-ForcePro+. Available online: http://www.oymotion.com/product17/82.

[38] Campbell E, Phinyomark A, Al-Timemy A H, et al. Differences in EMG feature space between able-bodied and amputee subjects for myoelectric control[C]//2019 9th International IEEE/EMBS Conference on Neural Engineering (NER). IEEE, 2019: 33-36.

[39] Chu J U, Moon I, Lee Y J, et al. A supervised feature-projection-based real-time EMG pattern recognition for multifunction myoelectric hand control[J]. IEEE/ASME Transactions on mechatronics, 2007, 12(3): 282-290.

[40] Xing K, Yang P, Huang J, et al. A real-time EMG pattern recognition method for virtual myoelectric hand control[J]. Neurocomputing, 2014, 136: 345-355.

[41] Altin C, Er O. Designing wearable joystick and performance comparison of EMG classification

methods for thumb finger gestures of joystick control[J]. Biomed Res India, 2017, 28: 4730-4736.
[42] Carvalho C R, Fernández J M, Del-Ama A J, et al. Review of electromyography onset detection methods for real-time control of robotic exoskeletons[J]. Journal of NeuroEngineering and Rehabilitation, 2023, 20(1): 141.
[43] Staude G, Wolf W. Objective motor response onset detection in surface myoelectric signals[J]. Medical engineering & physics, 1999, 21(6-7): 449-467.
[44] Navallas J, Ariz M, Villanueva A, et al. Optimizing interoperability between video-oculographic and electromyographic systems[J]. Journal of Rehabilitation Research & Development, 2011, 48(3).
[45] Ellaway P H. Cumulative sum technique and its application to the analysis of peristimulus time histograms[J]. Electroencephalography and clinical neurophysiology, 1978, 45(2): 302-304.
[46] Ghislieri M, Cerone G L, Knaflitz M, et al. Long short-term memory (LSTM) recurrent neural network for muscle activity detection[J]. Journal of NeuroEngineering and Rehabilitation, 2021, 18: 1-15.
[47] Morantes G, Fernández G, Altuve M. A threshold-based approach for muscle contraction detection from surface emg signals[C]//IX International Seminar on Medical Information Processing and Analysis. SPIE, 2013, 8922: 90-100.
[48] Staude G, Wolf W. Objective motor response onset detection in surface myoelectric signals[J]. Medical engineering & physics, 1999, 21(6-7): 449-467.
[49] Hodges P W, Bui B H. A comparison of computer-based methods for the determination of onset of muscle contraction using electromyography[J]. Electroencephalography and Clinical Neurophysiology/Electromyography and Motor Control, 1996, 101(6): 511-519.
[50] Izhikevich E M. Polychronization: computation with spikes[J]. Neural computation, 2006, 18(2): 245-282.
[51] Sun A, Chen X, Xu M, et al. Feasibility study on the application of a spiking neural network in myoelectric control systems[J]. Frontiers in Neuroscience, 2023, 17: 1174760.
[52] Xu M, Chen X, Sun A, et al. A Novel Event-Driven Spiking Convolutional Neural Network for Electromyography Pattern Recognition[J]. IEEE Transactions on Biomedical Engineering, 2023.
[53] Peng L, Hou Z G, Kasabov N, et al. Feasibility of NeuCube spiking neural network architecture for EMG pattern recognition[C]//2015 International Conference on Advanced Mechatronic Systems (ICAMechS). IEEE, 2015: 365-369.
[54] Ma Y, Chen B, Ren P, et al. EMG-based gestures classification using a mixed-signal neuromorphic processing system[J]. IEEE Journal on Emerging and Selected Topics in Circuits and Systems, 2020, 10(4): 578-587.
[55] Auge D, Hille J, Mueller E, et al. A survey of encoding techniques for signal processing in spiking neural networks[J]. Neural Processing Letters, 2021, 53(6): 4693-4710.
[56] Gerstner W, Kistler W M, Naud R, et al. Neuronal dynamics: From single neurons to networks and models of cognition[M]. Cambridge University Press, 2014.
[57] Kistler W M. Spike-timing dependent synaptic plasticity: a phenomenological framework[J]. Biological cybernetics, 2002, 87(5): 416-427.
[58] Wu Y, Deng L, Li G, et al. Spatio-temporal backpropagation for training high-performance spiking neural networks[J]. Frontiers in neuroscience, 2018, 12: 331.
[59] Zhang J, Liang M, Wei J, et al. A 28nm configurable asynchronous snn accelerator with energy-efficient learning[C]//2021 27th IEEE International Symposium on Asynchronous Circuits

and Systems (ASYNC). IEEE, 2021: 34-39.

[60] Horowitz M. 1.1 computing's energy problem (and what we can do about it)[C]//2014 IEEE international solid-state circuits conference digest of technical papers (ISSCC). IEEE, 2014: 10-14.

[61] Li X, Zhang X, Chen X, et al. Cross-user gesture recognition from sEMG signals using an optimal transport assisted student-teacher framework[J]. Computers in Biology and Medicine, 2023, 165: 107327.

[62] Cote-Allard U, Gagnon-Turcotte G, Phinyomark A, et al. A transferable adaptive domain adversarial neural network for virtual reality augmented EMG-based gesture recognition[J]. IEEE Transactions on Neural Systems and Rehabilitation Engineering, 2021, 29: 546-555.